ORIGINAL ARTICLE

# Novel Chern insulators with half-metallic edge states

Yang Xue[1], Bao Zhao[2], Yan Zhu[3], Tong Zhou[1], Jiayong Zhang[4], Ningbo Li[1], Hua Jiang[5] and Zhongqin Yang[1,6]

The central target of spintronics research is to achieve flexible control of highly efficient and spin-polarized electronic currents. Based on first-principles calculations and $k \cdot p$ models, we demonstrate that $Cu_2S$/MnSe heterostructures are a novel type of Chern insulators with half-metallic chiral edge states and a very high Fermi velocity ($0.87 \times 10^6$ m s$^{-1}$). The full spin-polarization of the edge states is found to be robust against the tuning of the chemical potential. Unlike the mechanisms reported previously, this heterostructure has quadratic bands with a normal band order, that is, the p/d-like band is below the s-like band. Charge transfer between the $Cu_2S$ moiety and the substrate results in variation in the occupied bands, which together with spin–orbit coupling, triggers the appearance of the topological state in the system. These results imply that numerous ordinary semiconductors with normal band order may convert into Chern insulators with half-metallic chiral edge states through this mechanism, providing a strategy to find a rich variety of materials for dissipationless, 100% spin-polarized and high-speed spintronic devices.

## INTRODUCTION

Microelectronic developments have given rise to an urgent need for new information storage and transport technologies and materials with very low energy consumption and high response speeds. Over the past two decades, great progress has been made in this regard in the following two areas. The first is based on the approach of spintronics by employing the spin degree of freedom of electrons that usually involves low power consumption. Typically, materials such as magnetic materials, half metals and spin-gapless semiconductors have been proposed for spintronics.[1–5] The second approach has benefited from the discovery of two-dimensional topological insulators, for which the edge states are expected to show dissipationless transport of the charge (or spin) current.[6–17] To date, few studies have combined both of these aspects well, namely, realizing dissipationless transport of a highly spin-polarized current through the chiral edge states of materials such as Chern insulators.[18] It has been reported that the edge states of Chern insulators are generally not spin-polarized[19] or are merely partially spin-polarized in the sample plane.[20–22] Therefore, it is extremely desirable to realize dissipationless electronic transport through the Chern insulators with half-metallic chiral edge states, offering great potential for applications of the chiral edge states of Chern insulators in spintronics.

Proposals to produce the Chern insulators generally exploit Dirac electrons[23] and employ the mechanism of special inverted bands.[24] These Chern insulators are generally limited to small nontrivial band gaps, show problematic disorder due to magnetic doping, and are found at very low experimental temperatures.[24] To overcome these drawbacks of Dirac electron systems, one may explore the topological states in numerous ordinary semiconductor materials, for which the dispersions are usually quadratic and show normal band order. Actually, the quantum spin Hall effect has been reported in several non-Dirac electron systems.[25–31] The appearance of the quantum spin Hall state in these systems is associated with the inverted band order, that is, the p/d-like bands are abnormally above the s-like band (Figure 1a),[25–30] which has been regarded as an essential and effective strategy for the generation of nontrivial topological states.[24] Nevertheless, perhaps it is possible to realize nontrivial topology effects in non-Dirac electron systems with the normal band order that is found in many semiconductor materials such as Ge, GaAs, silicene, and germanene.[32,33] A straightforward approach for achieving a topological state from a non-Dirac dispersion with normal band order is illustrated in Figures 1b and c. The bands displayed in Figure 1b give the typical energy dispersion relationship around the Fermi level ($E_F$) of non-Dirac-type semiconductor materials with normal band order. If the bands shown in Figure 1c can be achieved, a nontrivial band gap may be opened in Figure 1c around the $E_F$ after the intrinsic spin–orbit coupling (SOC) is considered. If both branches of the p/d-like bands are in a single spinchannel (spin-up or spin-down), the edge states generated within the nontrivial band gap may be fully spin-polarized.

In this paper, we report a novel type of Chern insulators with half-metallic (100% spin-polarized) chiral edge states in the β-$Cu_2S$ films grown on MnSe(111) surfaces. The thin film of β-$Cu_2$S is an ordinary non-Dirac-type semiconductor with a normal band order, as shown in Figure 1b, which has been synthesized experimentally and has been found to be stable at room temperature.[34,35] The MnSe substrate provides magnetization in the β-$Cu_2$S bands to break the time-reversal symmetry. Based on density-functional theory and the Wannier function method, we find that these novel Chern insulators exhibit several prominent advantages: (i) The chiral edge states are fully spin-polarized, and the spin polarization ratio does not change with variation of the chemical potential. (ii) A very high Fermi velocity ($0.87 \times 10^6$ m s$^{-1}$) is achieved, comparable to that of graphene. (iii) This type of Chern insulator is achieved in non-Dirac electrons with normal band order, implying that many ordinary semiconductor materials may become Chern insulators through this mechanism. (iv) The nontrivial band gap is sizeable (45 meV), supporting the room-temperature experiments and applications of the Chern insulating behaviors. These discoveries create a new path for the design of Chern insulators with half-metallic chiral edge states from ordinary semiconductor materials and provide new opportunities to realize highly efficient spintronic devices.

[1]State Key Laboratory of Surface Physics and Key Laboratory for Computational Physical Sciences (MOE) and Department of Physics, Fudan University, Shanghai, China; [2]College of Physical Science and Information Technology, Liaocheng University, Liaocheng, China; [3]College of Science, Nanjing University of Aeronautics and Astronautics, Nanjing, China; [4]Jiangsu Key Laboratory of Micro and Nano Heat Fluid Flow Technology and Energy Application, School of Mathematics and Physics, Suzhou University of Science and Technology, Suzhou, Jiangsu, China; [5]College of Physics, Optoelectronics, and Energy, Soochow University, Suzhou, China and [6]Collaborative Innovation Center of Advanced Microstructures, Fudan University, Shanghai, China
Correspondence: Professor ZQ Yang, Department of Physics, Fudan University, 220 Handan Rd, Shanghai 200433, China.
E-mail: zyang@fudan.edu.cn
Received 16 August 2017; revised 8 November 2017; accepted 26 November 2017

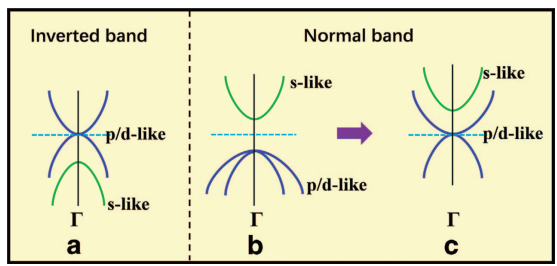

**Figure 1** Schematic diagram of the inverted bands (**a**), normal bands (**b**), and tuned bands with normal band order (**c**). The blue and green solid curves indicate the p/d-like bands and s-like bands, respectively. The purple arrow represents the tuning process of the bands with normal band order without the consideration of SOC. The blue dashed lines indicate the Fermi level. SOC, spin–orbit coupling.

## MATERIALS AND METHODS

The constructed heterostructure of the β-$Cu_2$S monolayer on MnSe(111) surfaces is displayed in Figures 2a and b. Very thin β-$Cu_2$S films have been fabricated experimentally.[34,35] One exotic characteristic of this kind of $Cu_2$S films (Figure 2c) is that the structure usually exists in a solid–liquid hybrid phase with a framework of graphene-like fixed S layers and mobile (liquid-like) Cu atoms within the framework.[34–36] For the substrate material, the bulk MnSe compound has a cubic rocksalt structure in which the Mn atoms in the (111) plane form a ferromagnetic (FM) triangular lattice, and the neighboring two Mn(111) planes, containing one Se plane between them, form an antiferromagnetic structure.[37,38] The $Cu_2$S monolayer can be magnetized by the topmost FM Mn plane through the proximity effect.[13,14] The lattice of the β-$Cu_2$S monolayer[35] (Figure 2c) matches well with that of the MnSe (111) plane (with a small mismatch of 2%). The structural details and different contact configurations considered for the $Cu_2$S and MnSe interfaces can be found in the Supporting Information. The structure displayed in Figures 2a and b is found to be the most stable configuration (top site), and its electronic structures and topology will be discussed in detail below. A relatively strong chemical bonding exists between the bottom Cu atoms and the topmost Mn atoms, confirmed by the large value of the binding energy (2 eV per unit cell). This tendency is consistent with the small distance $d$ ($d = 1.39$ Å) between the bottom Cu plane and the topmost Mn layer in the substrate after the geometry optimization (see Figure 2a). The structural stability is also further confirmed by molecular dynamics simulations (see Supporting Information).

The geometric optimization and electronic structure calculations of the $Cu_2$S monolayer grown on the MnSe substrate are performed with the projector augmented wave method based on *ab initio* density-functional theory.[39,40] The Perdew–Burke–Ernzerhof generalized gradient approximation (GGA) is used to describe the exchange and correlation functional.[41] The Mn 3d orbitals are treated with the GGA plus Hubbard U (GGA+U) method,[42] in which the effective on-site Coulomb interaction $U$ and exchange interaction $J$ are set to 5.0 and 1.0 eV,[43,44] respectively. The plane-wave cutoff energy is set to 500 eV, and a vacuum space larger than 15 Å is adopted to avoid the interaction between the two adjacent slabs. The $9 \times 9 \times 1$ gamma central Monkhorst–Pack grids are employed to perform the first Brillouin zone (BZ) integral. MD simulations are carried out to study the structural stability using Born–Oppenheimer ground states with 4 fs as the time step. A thermostat is used to control the temperature to remain at room temperature (300 K). The structure has been simulated for 2.4 ps. The edge state and its spin-texture are calculated by using the surface Green's function method.[45]

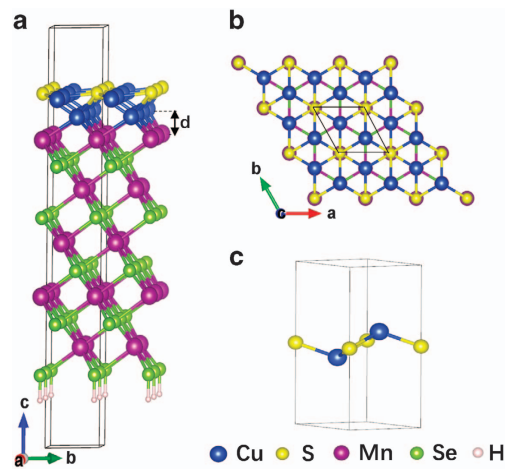

**Figure 2** Side (**a**) and top (**b**) views of the β-$Cu_2$S monolayer grown on the MnSe (111) surface. The black thin lines in (**b**) show the unit cell. (**c**) Side view of the β-$Cu_2$S monolayer. The Cu, S, Mn, Se, and H atoms are denoted in blue, yellow, purple, green, and light gray, respectively.

## RESULTS AND DISCUSSION

We now explore the topological electronic structures of the $Cu_2$S/MnSe heterostructure shown in Figures 2a and b. The band structures without and with consideration of SOC are given in Figures 3a and b, respectively. As shown in Figure 3a, when the $Cu_2$S monolayer is placed on the (111) surface of the MnSe substrate, an obvious spin splitting appears in the $Cu_2$S bands. The $Cu_2$S is magnetized well by the topmost FM Mn layer. Our calculations also show that the magnetization of the heterostructure along the c-axis is 4 meV more stable than that lying in the *ab* plane, enabling improvements in the experimental observation of the quantum anomalous Hall (QAH) effect. Without the SOC, the $Cu_2$S/MnSe system is gapless, with the top of the valence bands and the bottom of the conduction bands degenerate at the Γ point in the spin-up channel, as displayed in Figure 3a. The degeneracy is primarily composed of Cu $d_{x^2-y^2}$ and $d_{xy}$ orbitals and arises from the $C_{3v}$ symmetry of the heterostructure. However, the SOC splits the double degeneracy at the Γ point, resulting in a global and sizeable band gap of approximately 45 meV opening around the $E_F$ (Figure 3b), which is large enough to support the corresponding experimental work and possible applications of the heterostructure at room temperature. To identify the topological nature of the insulating phase, the Chern number is calculated by integrating the Berry curvatures over the first BZ.[46–50] The red curve in Figure 3c gives the obtained Berry curvatures along the high-symmetry lines in the first BZ. The obtained Chern number of $C=-1$ demonstrates the nontrivial topological features of the band gap opened by the SOC.

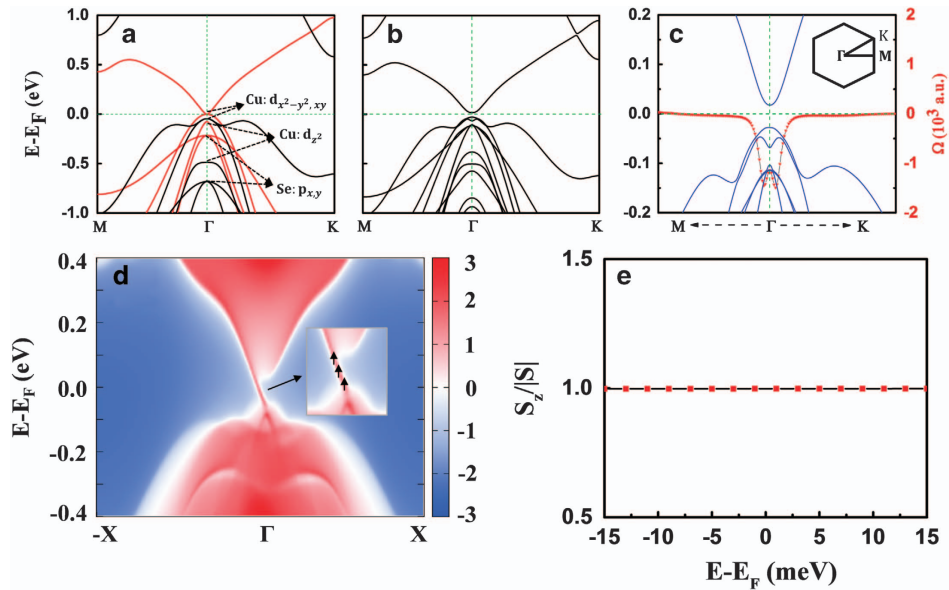

**Figure 3** (**a**, **b**) Band structures of the Cu$_2$S/MnSe heterostructure without and with SOC considered, respectively. The red and black curves in (**a**) denote the spin-up and spin-down states, respectively. (**c**) The calculated Berry curvature (red) for all valance bands. The blue curves represent the bands obtained from the Wannier functions. The inset shows the first BZ of the system with the high-symmetry points. (**d**) The topological protected edge state is obtained at the edge of the semi-infinite plane of the slab system. The inset shows the spin direction of the edge state (along the *c*-axis). (**e**) The spin texture of the edge state within the bulk band gap. The *y*-axis gives the *z*-component (along the *c*-axis) of the spin divided by the spin modulus. SOC, spin–orbit coupling.

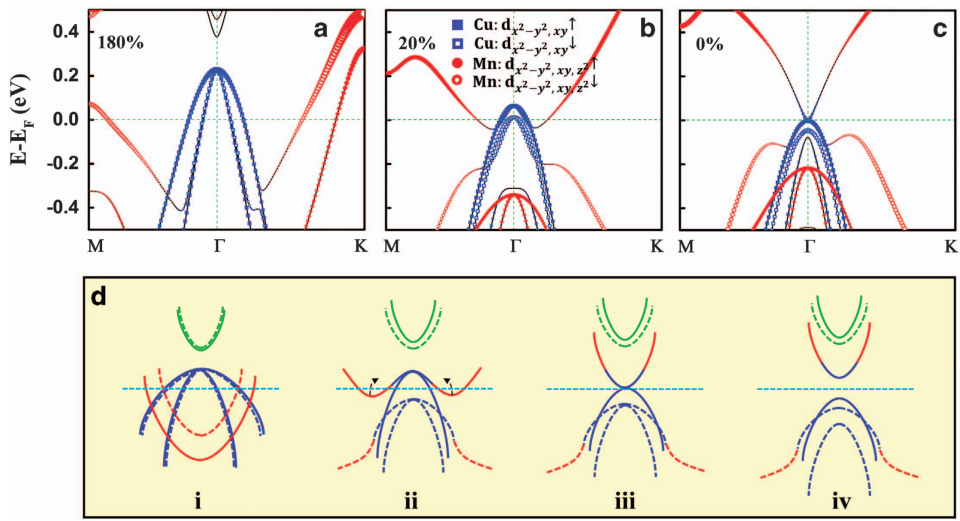

**Figure 4** Orbital-resolved band structures of the Cu$_2$S/MnSe heterostructure without SOC under tensile strain of (**a**) 180%, (**b**) 20%, and (**c**) 0%. The components of the Mn d states are multiplied by a factor of 2.5 for clarity. (**d**) The first three cases are the schematic diagram of the band evolution corresponding to (**a**–**c**). The last case (iv) gives the bands of (iii) with SOC. The green curves in (**d**) denote the Cu s-like bands. The black arrows in (ii) indicate the electron transfer from the Mn d bands to the Cu d bands. The blue dashed lines in (**d**) indicate the $E_F$. SOC, spin–orbit coupling.

The obtained edge state (red curve) of the heterostructure is displayed in Figure 3d and is located at the edge of the semi-infinite plane of the slab system. In particular, the calculated spin texture of the edge state (Figure 3e) shows that the edge state is 100% spin-polarized in the *z*-direction (along the *c*-axis), indicating the half-metallic behavior of the chiral edge state (see the inset of Figure 3d). The half-metallic behavior obtained here is distinct from the traditional half-metallic property[3] for which the spin-polarization may be highly suppressed upon consideration of the SOC.[51] Importantly, the constant value of $S_z/|S|$ in Figure 3e demonstrates that the full spin-polarization is robust against the tuning of the chemical potential, unlike the case in Zhang et al.[22]; this is beneficial for the applications of the edge states in spintronics. The one-dimensional (1D) boundary of the heterostructures slab is a closed manifold in real space, and the electrons in the chiral edge state will acquire a quantized π Berry phase after they evolve one cycle along the boundary. Thus, the full spin-polarization of the edge state can be recognized as the 1D real-space topological states.[20] The half-metallic behavior of the edge state can be associated with the bulk nontrivial band gap introduced by both bands in the spin-up channel (Figure 3a) without SOC. The conservation of the $S_z$ of the edge states (Figures 3d and e) indicates that the Rashba SOC plays a negligible role in the band-gap opening. Namely, the band gap is primarily opened by atomic SOC. The sharp slope of the almost linear dispersion of the edge state implies the very high Fermi velocity $v_F$ (~$0.87 \times 10^6$ m s$^{-1}$) of the electrons, as evaluated by ($\hbar v_F \approx \partial E(k)/\partial k$). This Fermi velocity is even comparable to that of graphene ($1 \times 10^6$ m s$^{-1}$).

These results prove that the Cu$_2$S/MnSe heterostructure is a novel Chern insulator with exotic half-metallic chiral edge states. Because of these edge states, the Cu$_2$S/MnSe heterostructure has great potential for the fabrication of a dissipationless, 100% spin-polarized, and high-speed spintronic devices and is superior to the Chern insulators with no spin polarization or partially spin-polarized edge states.[19,21,22] It is also important to note that the $E_F$ in this system lies in the nontrivial band gap, enabling improvements in the experimental observations.

To understand the origin of the Chern insulators, the band evolution of the heterostructure with different contact distances is explored. Since the influence of the MnSe substrate on the Cu$_2$S sample can be tuned by changing the distance ($d$) between the bottom Cu plane and the topmost Mn plane (Figure 2a), how the substrate varies the bands of the Cu$_2$S film can be examined by tuning the distance $d$. The projected band structures with different contact distances $d$ are displayed in Figures 4a–c, where the blue squares and red circles represent the Cu d$_{x^2-y^2}$/d$_{xy}$ and Mn d$_{x^2-y^2}$/d$_{xy}$/d$_{z^2}$ orbitals, respectively. SOC is not considered in this figure. When imposing a 180% tensile strain (defined as $(d-d_0)/d_0$, where $d$ stands for the distance between the bottom Cu plane and the topmost Mn plane with the applied strain and $d_0$ stands for the relaxed distance without the strain (1.39 Å)), the Cu bands (blue curves) cross the Mn bands (red curves). Importantly, no spin splitting is found for the Cu bands (blue curves) in Figure 4a because of the very weak interaction from the topmost FM Mn plane in the substrate ($d = 3.89$ Å).

Upon decreasing the tensile strain from 180% to 20%, the Cu bands hybridize strongly with the Mn bands around the $E_F$, as illustrated in Figure 4b. The spin polarization of the Cu d bands is induced by the Mn plane due to the chemical bonding formed between the interface Cu and Mn atoms. In particular, the Cu bands move downward in energy while the Mn bands show the opposite effect. Thus, electron transfer occurs from the Mn atoms in the substrate to the Cu atoms in the Cu$_2$S sample, which is likely due to the higher electronegativity of the Cu atoms relative to that of the Mn atoms. With the tensile strain further decreasing to zero (Figure 4c), more Mn d electrons transfer to the Cu d states. Since the stretched strain occurs only in the c direction (Figure 2a), the $C_{3v}$ symmetry is not broken in the process. Hence, the double degeneracy of the Cu d orbitals (d$_{x^2-y^2}$, d$_{xy}$) at the Γ point is preserved in the process of the tensile strain variation (Figures 4a and b). A gapless semiconductor is finally achieved in Figure 4c without consideration of SOC, with the $E_F$ crossing exactly the degenerate point of Cu d$_{x^2-y^2}$ and d$_{xy}$.

The schematic diagram of the above band evolution is illustrated in Figure 4d, where the solid and dashed curves represent the spin-up and spin-down bands, respectively. The green, blue, and red curves denote the Cu$_2$S s-like bands, d-like bands and MnSe d-like bands, respectively. As seen in stage (i), when the distance between the monolayer Cu$_2$S and the substrate is relatively large, no exchange field is induced in the Cu$_2$S bands, which has a typical semiconducting band structure with the normal band order. With a decrease in the distance between the Cu$_2$S film and the substrate, the exchange field, chemical bonding, and charge transfer (indicated by the black arrows) finally give rise to a gapless semiconductor (from stage (ii) to (iii)). After SOC is included, the quadratic band touching at the Γ point is no longer protected by the $C_{3v}$ symmetry according to the double group representations of $C_{3v}$ symmetry.[52] Thus, a sizable nontrivial band gap is acquired, and a QAH insulator is obtained in the Cu$_2$S/MnSe heterostructure (stage (iv)).

It is important to emphasize that the chemical bonding and charge transfer at the interface of the sample and the substrate result in band deformation (Figures 1b and c) and lead to the QAH state in the heterostructure. The band evolution in Figure 4d clearly illustrates that the formation of the topological states does not at all demand the band inversion of the s-like band below the p/d-like band, differing from many previously reported cases in non-Dirac electron systems.[25–30] Although the s-band in Figures 1b, c and 4 is not essential for the appearance of the topological state in the Cu$_2$S/MnSe heterostructure, it is displayed for a good comparison with the traditional inverted band mechanism of the topological states proposed in numerous non-Dirac systems (Figure 1a). Furthermore, many semiconductor materials show normal band ordering around the Γ point, that is, the p/d-like band is below the s-like band. Hence, our work offers more accessible materials for experimental observation and applications of the Chern insulators. Since no band inversion is required, the exchange interaction does not need to be very large to generate the QAH effect, as seen from the last two stages in Figure 4d. Moreover, the nontrivial gap is opened by the intrinsic atomic SOC rather than by the external Rashba SOC, resulting in a rather large band gap (that is, approaching the atomic SOC strength). Our results show that many non-Dirac-type semiconductor materials with normal band order may produce the QAH effect based on the mechanism proposed here, providing a new route for discovery of the Chern insulators.

To deeply understand the topological mechanism, an effective Hamiltonian was then constructed in terms of the k·p model for low-energy physics and based on the invariant theory.[31,53] Since the bands near the $E_F$ around the Γ point are dominated by d$_{x^2-y^2}$ and d$_{xy}$ orbitals of the Cu atoms in the Cu$_2$S monolayer, it is reasonable to adopt these two orbitals as the basis. For convenience, the two orbitals are transformed to the basis of $\{|d_{+,2\uparrow}\rangle, |d_{-,2\uparrow}\rangle, |d_{+,2\downarrow}\rangle, |d_{-,2\downarrow}\rangle\}$ with $|d_{\pm,2}\rangle = (1/\sqrt{2})(|d_{x^2-y^2}\rangle \pm i|d_{xy}\rangle)$ and ↑(↓) for the spin-up (-down) state. The Hamiltonian takes the form:[31]

$$H_0 = \begin{pmatrix} \epsilon_1(k) & f(k) & 0 & 0 \\ f^*(k) & \epsilon_2(k) & 0 & 0 \\ 0 & 0 & \epsilon_1(k) & f(k) \\ 0 & 0 & f^*(k) & \epsilon_2(k) \end{pmatrix} \quad (2)$$

In the absence of SOC and magnetism, the system contains time-reversal symmetry (T), mirror symmetry (M$_l$) along the $l$-axis ($\vec{l} = \vec{a} + (1/2)\vec{b}$) direction, and threefold rotation ($C_3$) along the c-axis. Owing to the T symmetry, $f(k)$ should be an even function of k. By taking the $C_3$ symmetry into consideration, one can obtain $f(k_\pm) = e^{i\frac{2\pi}{3} \times 4} f(e^{\pm i\frac{2\pi}{3}} k_\pm)$, where $k_\pm = k_x \pm ik_y$ because the angular momentum of the d$_{x^2-y^2}$ and d$_{xy}$ orbitals equals 2. Thus, by inspecting the two constraints of T and $C_3$, we see that $f(k)$ must take the form of $f(k) = \beta k_+^2$. Moreover, we have $\epsilon_1(k) = \epsilon_2(k) = \alpha k^2$, $k = |k|$. The FM exchange term as $H_M = M\sigma_z \otimes 1$ ($\sigma_z$ is the Pauli matrix) can be added to the Hamiltonian (Equation (2)). Such a term can separate the energy bands for spin-up and spin-down states, as found by the density-functional theory calculations. Because the Rashba SOC is trivial compared to the exchange interaction, we can neglect the coupling between the spin-up and spin-down states and proceed to the discussion of the spin-up channel, that is, on the basis of $\{|d_{+,2\uparrow}\rangle, |d_{-,2\uparrow}\rangle\}$.

For the atomic SOC effect, $H_{so} = \lambda_{so} L \cdot S$ is diagonal in the selected basis. Thus, the k·p model under $\{|d_{+,2\uparrow}\rangle, |d_{-,2\uparrow}\rangle\}$ around the Γ point

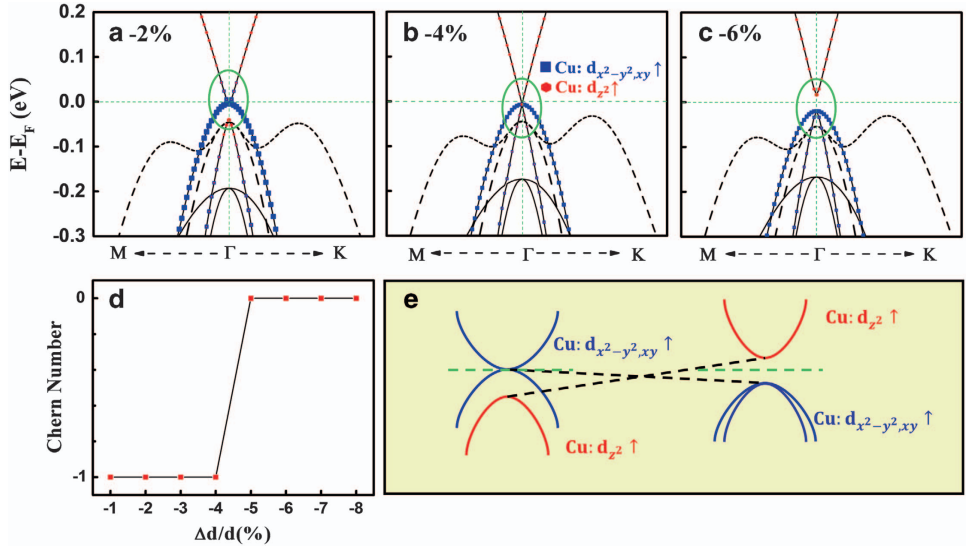

**Figure 5** Band structures of theCu$_2$S/MnSe heterostructure without SOC under compressive strain in the *c*-axis direction of (**a**) − 2%, (**b**) − 4%, and (**c**) − 6%. The components of the Cu spin-up d$_{x^2-y^2}$/d$_{xy}$ and d$_{z^2}$ orbitals are indicated in (**a–c**), where the solid and dashed black curves denote the spin-up and spin-down states, respectively. The component of the Cu spin-up d$_{z^2}$ orbital is multiplied by a factor of 3 for clarity. The green ovals in (**a–c**) indicate the regions interested. (**d**) Strain-induced evolution of the calculated Chern numbers (red points) for the whole valence bands with consideration of SOC. (**e**) Schematic diagram of the band evolution of the heterostructure under the compressive process. The green dashed lines in (**e**) indicate the $E_F$. SOC, spin–orbit coupling.

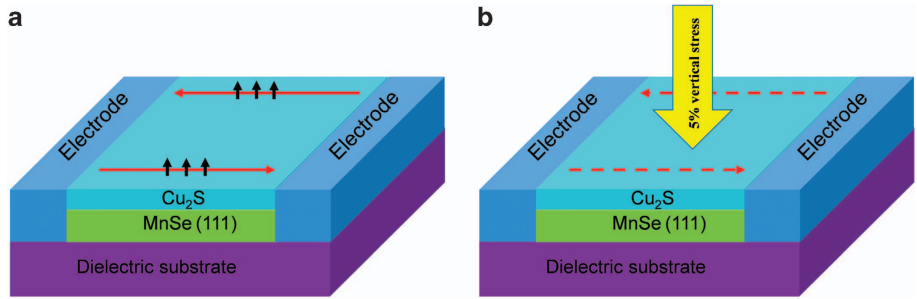

**Figure 6** (**a**) The designed dissipationless and high-speed spintronic device. The heterostructure is deposited on a dielectric substrate and sandwiched between two electrodes, which can drive net electric current conducting through the edge states in the sample. The fully spin-polarized charge current (red arrows) is conducted through the topologically nontrivial half-metallic edge states in the sample. The small black arrows indicate the spin direction, perpendicular to the sample plane. (**b**) The device can be switched off by applying a vertical stress. The edge states disappear when the vertical stress is equal to or larger than 5%.

can be constructed up to the second order of *k* according to:

$$H = (\alpha k^2 + M) \otimes 1 + \begin{pmatrix} \lambda_{so} & \beta k_+^2 \\ \beta^* k_-^2 & -\lambda_{so} \end{pmatrix} \quad (4)$$

To explore the topology and pseudospin texture, we neglect the constant term $M \otimes 1$, which does not modify the eigenstates of the system. Thus, the Hamiltonian *H* is rewritten as:

$$H_0 = \alpha k^2 \otimes 1 + d \cdot \sigma \quad (5)$$

with the $d = (d_x, d_y, d_z) = \left(\beta(k_x^2 - k_y^2), -2\beta k_x k_y, \lambda_{so}\right)$. The calculated pseudospin texture $\hat{d} = \frac{(\{\beta(k_x^2 - k_y^2), -2\beta k_x k_y, \lambda_{so}\})}{(\sqrt{\beta^2 k^4 + \lambda_{so}^2})}$ for the total Hamiltonian is shown in Supplementary Figure S2a. An obviously very special two-meron structure with double vorticities is observed in the pseudospin texture.[6,31] By calculating the pseudospin Chern number as $n = 1/4\pi \int dk^2 (\partial_{k_x}\hat{d} \times \partial_{k_y}\hat{d})\cdot\hat{d}$, the Chern number of − 1 is acquired, in agreement with the result obtained from the density-functional theory calculations described above and with the result acquired by integrating the following Berry curvature:

$$\Omega(k) = \frac{-2\beta^2 \lambda_{so} k^2}{\sqrt{(\beta^2 k^4 + \lambda_{so}^2)^3}} \quad (7)$$

derived from $\Omega(k) = (d/d^3)\cdot(\partial d/\partial k)$.[23] It is interesting to find that the systems with non-Dirac band dispersions composed of other orbitals such as p$_{x,y}$ or d$_{xz,yz}$, have Chern numbers with opposite sign and inverse edge states for the same spin channel due to their different angular momentum. The corresponding pseudospin textures also have the opposite chirality (Supplementary Figure S2c). Thus, the special two-meron pseudospin texture shown in Supplementary Figures S2a and b is the origin of the QAH effect in the non-Dirac electron system of the Cu$_2$S/MnSe heterostructure, with the normal band order. The result shown in Supplementary Figure S2c may be found in other non-Dirac electron material systems containing interesting degenerate bands composed of p$_{x,y}$ or d$_{xz,yz}$ instead of d$_{x^2-y^2, xy}$.

The exchange field in the monolayer Cu$_2$S is needed for the QAH effect and is sensitive to the interface distance (*d*) between the monolayer and the substrate (Figure 2a). Here, we reveal the effect observed when the exchange field is enhanced by decreasing the interface distance ($d$) from −1% to −8% along the $c$-axis, which may be realized experimentally by applying an external vertical stress to the heterostructure.

Figures 5a–c illustrate the band structures without SOC under various compressive strains. Under a compressive strain of −2%, the Cu spin-up $d_{z^2}$ band close to the Γ point is at about −0.05 eV, denoted by the red hexagons (Figure 5a). By increasing the compressive strain to −4%, the $d_{z^2}$ band moves up drastically in energy and touches the degenerate point of $d_{x^2-y^2, xy}$ (the blue squares) at the Γ point (Figure 5b). When increasing the compressive strain to −6%, the Cu spin-up $d_{z^2}$ band inverts with one of the branches of the degenerate bands (the blue squares), and a gap appears between the degenerate bands and the inverted $d_{z^2}$ band (Figure 5c). The Chern number as a function of the compressive strains is calculated and plotted in Figure 5d, in which there is a jump of Chern number from −1 to 0 at the compressive strain of −4%. This indicates a topological phase transition under the compressive strain, that is, the QAH effect in the heterostructure will be destroyed by compressive strains larger than −4%. The origin of the Cu spin-up $d_{z^2}$ band energy upshift can be ascribed to the stronger spin polarization of the Cu $d_{z^2}$ band with the increase of the compressive strain in the $c$-axis direction. Thus, the exchange field becomes large, with the Cu spin-up $d_{z^2}$ moving up and spin-down $d_{z^2}$ shifting down in energy. By analyzing the band components, the mechanism of the topological phase transition is displayed in Figure 5e, in which the green dashed lines represent the $E_F$. Because the degenerate bands of Cu $d_{x^2-y^2, xy\uparrow}$ invert with the Cu $d_{z^2\uparrow}$ band, both degenerate Cu $d_{x^2-y^2, xy\uparrow}$ bands around the Γ point become occupied. Additionally, a trivial gap is generated around the $E_F$. Our results suggest that the novel Chern insulator predicted in the β-Cu$_2$S/MnSe heterostructure can survive very well with small compressive strains.

Based on the achieved half-metallic nontrivial edge states, the dissipationless, 100% spin-polarized, and high-speed electronic transport can be realized in a single device, as shown in Figure 6a. Based on the phase transition given in Figure 5d, we predict that the on–off of this device can be tuned effectively by applying a small vertical stress (Figure 6b) because such stress can make the nontrivial edge states of the sample disappear.

In summary, we proposed a novel Chern insulator with unique half-metallic chiral edge states in the β-Cu$_2$S/MnSe heterostructures. This type of Chern insulator is achieved in non-Dirac electrons with a normal band order, implying that many ordinary semiconductors may become Chern insulators via this mechanism. The interface charge transfer is found to be important for giving rise to the topological behavior. The robust nontrivial edge state is found to be 100% spin-polarized and to have a very high Fermi velocity ($0.87 \times 10^6$ m s$^{-1}$), close to that of graphene. The $k \cdot p$ model shows that the topology in the system originates from the exotic two-meron structure of the pesudospin texture. These results indicate that the β-Cu$_2$S/MnSe heterostructure has great potential to be fabricated into an advanced dissipationless, 100% spin polarized, and high-speed spintronic device. Our work also creates a rational path to design novel Chern insulators with very special half-metallic chiral edge states.

## CONFLICT OF INTEREST
The authors declare no conflict of interest.

## ACKNOWLEDGEMENTS
The authors are grateful to Dr Quansheng Wu for very helpful discussion. This work was supported by the National Natural Science Foundation of China under Grant No. 11574051 and the Natural Science Foundation of Jiangsu Province of China (Grant No. BK20160007). All calculations were performed at the High-Performance Computational Center (HPCC) of Department of Physics at Fudan University.

## PUBLISHER'S NOTE
Springer Nature remains neutral with regard to jurisdictional claims in published maps and institutional affiliations.